\documentclass{PoS}

\usepackage{amssymb}
\usepackage{fontenc}
\usepackage{times}
\usepackage{mathptmx}

\newcommand{\mbar}{\overline{m}}
\newcommand{\muh}{\delta {m}_u}
\newcommand{\mdh}{\delta {m}_d}
\newcommand{\msh}{\delta {m}_s}
\newcommand{\mlh}{\delta {m}_l}

\newcommand{\dml}{\delta {m}_l}

\newcommand{\tophat}{$\phantom{I^{I^{I^I}}}\hspace{-6.5mm} $}

\newcommand{\ts} {\textstyle}

\newcommand{\R}{\mbox{\tiny $R$}}                        

\title{
\vspace*{-1.25cm}
\begin{minipage}{\textwidth}
\begin{flushright}
\texttt{\footnotesize
PoS(Lattice 2012)116 \\
ADP-12-50/T817       \\
DESY 12-229          \\
Edinburgh 2012/23    \\
Liverpool LTH 967    \\
}
\end{flushright}
\end{minipage}\\[15pt]
\vspace*{+1.25cm}
       The effects of flavour symmetry breaking on hadron matrix
       elements}

\ShortTitle{Flavour symmetry breaking $\ldots$
            \hspace*{2.425in}\rm{A.~N. Cooke and P.~E.~L. Rakow}}

\author{A.~N. Cooke\thanks{Joint speakers}\footnotemark[1]$\,\,\,^{\,a}$,
        R. Horsley$^{a}$,
        Y. Nakamura$^b$,
        D. Pleiter$^{cd}$, 
        P.~E.~L. Rakow\footnotemark[1]$\,\,\,^e$,
        G.~Schierholz$^f$
        and J.~M. Zanotti$^{g}$  \\
        \llap{$^a$} School of Physics and Astronomy,
                    University of Edinburgh,
                    Edinburgh EH9 3JZ, UK \\
        \llap{$^b$} RIKEN Advanced Institute for Computational Science,
                    Kobe, Hyogo 650-0047, Japan \\
        \llap{$^c$} JSC, J\"ulich Research Centre,
                    52425 J\"ulich, Germany \\
        \llap{$^d$} Institut f\"ur Theoretische Physik,
                    Universit\"at Regensburg, 93040 Regensburg, Germany \\
        \llap{$^e$} Theoretical Physics Division,
                    Department of Mathematical Sciences,
                    University of Liverpool,
                    Liverpool L69 3BX, UK \\
        \llap{$^f$} Deutsches Elektronen-Synchrotron DESY,
                    22603 Hamburg, Germany \\
        \llap{$^g$} CSSM, School of Chemistry and Physics,
                    University of Adelaide, Adelaide SA 5005, Australia \\
        E-mail: \email{ashley.cooke@ed.ac.uk}, \email{rakow@amtp.liv.ac.uk}}

\abstract{
 By considering a flavour expansion about the $SU(3)$-flavour symmetric 
 point, we investigate how flavour-blindness constrains octet baryon 
 matrix elements after $SU(3)$ is broken by the mass difference between 
 the strange and light quarks. We find the expansions to be highly 
 constrained along a mass trajectory where the singlet quark mass is 
 held constant, which proves beneficial for extrapolations of 2+1 flavour 
 lattice data to the physical point. We investigate these effects 
 numerically via a lattice calculation of the flavour-conserving and 
 flavour-changing matrix elements of the vector and axial operators 
 between octet baryon states.}

\FullConference{The 30th International Symposium on Lattice Field Theory
                - Lattice 2012,\\
		June 24-29, 2012\\
		Cairns, Australia}

\begin{document}


\section{Introduction}


Understanding the pattern of flavour symmetry breaking and mixing, and
the origin of CP violation, remains one of the outstanding problems in
particle physics. In \cite{bietenholz10a,bietenholz11a}
we have outlined a program to systematically investigate the pattern
of flavour symmetry breaking. The program has been successfully applied
to meson and baryon masses involving up, down and strange quarks.
In these talks we will extend the investigations to include matrix elements.

The QCD interaction is flavour-blind. Neglecting
electromagnetic and weak interactions, the only difference
between flavours comes from the mass matrix. We investigate
how flavour-blindness constrains matrix elements after flavour
$SU(3)$ is broken by the mass difference between the strange and
light quarks, to help us extrapolate $2+1$ flavour lattice data
to the physical point.

We have our best theoretical understanding when all 3 quark flavours
have the same masses (because we can use the full power of flavour
$SU(3)$); nature presents us with just one instance of the theory,
with $ m_s/m_l \approx 25.$ On the lattice we can choose our quark masses,
so we can investigate fictional universes where $ m_s / m_l \ne 25$,
and so gain a clearer understanding of flavour symmetry breaking.

We have previously used a symmetry analysis, similar in spirit
to that used by Gell-Mann and Okubo \cite{gell-mann62a,okubo62a}
in the earliest days of the quark picture, to find formulae for
the quark mass dependence of hadron masses \cite{bietenholz10a,bietenholz11a}.
We now extend this analysis to hadron matrix  elements. In the first
part of these proceedings we discuss the group theory, in the second
part we compare our expectations with lattice data. We shall then
discuss briefly the further steps needed to compute form factors
relevant to the determination of the CKM matrix element $|V_{us}|$.

In this work we concentrate on the $2+1$ case, in which symmetry
breaking is due to mass differences between the strange and light
quarks; but our methods are also applicable to isospin breaking 
effects coming from a non-zero $m_d-m_u$, e.g.\ \cite{horsley12a}.


\section{$SU(3)$ breaking}
\label{su3_breaking}


How severely does the strange quark mass break 
$SU(3)$ symmetry? In this approach it is not the strange-light ratio,
$ m_s / m_l \sim 25 $, which matters. A more natural way to
judge the severity of symmetry breaking is to compare $(m_s - m_l)$
with a typical hadronic mass. Since $(m_s - m_l) \ll M_B$ 
we can hope for an expansion with very good convergence. 

We can see how well this works in practice by looking,
for example, at the physical masses of the decuplet baryons.
We can construct mass combinations which first appear at 
different orders of the symmetry breaking, starting with 
quantities that would be non-zero even with perfect $SU(3)$, 
and working up to quantities which first appear at the third
order in the symmetry breaking parameter
\begin{eqnarray}
   4 M_\Delta + 3 M_{\Sigma^*} + 2 M_{\Xi^*} + M_{\Omega}
      &=& +13.82 {\rm \ GeV}
          \qquad \quad \ {\rm singlet}
          \quad \propto (m_s- m_l)^0 \label{num1}          \nonumber  \\
   - 2 M_\Delta \qquad\ \quad  + M_{\Xi^*} + M_{\Omega}
      &=& +0.742 {\rm \ GeV}
          \qquad \quad \ {\rm octet}
          \quad\,\,\,\,\,\, \propto (m_s- m_l)^1 \label{num8} \nonumber  \\
   4 M_\Delta - 5 M_{\Sigma^*} - 2 M_{\Xi^*} + 3 M_{\Omega}
      &=& -0.044  {\rm \ GeV}
          \qquad \quad \mbox{27-plet}
          \quad \propto (m_s- m_l)^2 \label{num27}         \nonumber  \\
   - M_\Delta + 3 M_{\Sigma^*} - 3 M_{\Xi^*} +  M_{\Omega}
      &=& -0.006 {\rm \  GeV}  \qquad \quad \mbox{64-plet} \quad
          \propto  (m_s- m_l)^3\,,
\label{num64}
\end{eqnarray}
where we have used the isospin-averaged experimental masses for 
each decuplet baryon. Clearly we see a strong hierarchy in values. 
Each additional factor of $(m_s- m_l)$ reduces the value by about 
an order of magnitude, the final $O((m_s- m_l)^3 )$ quantity is 
about $2000$ times smaller than the leading quantity. An expansion
that yields a factor of $10$ for each order is very good compared
with most approaches available for QCD. 

To investigate flavour symmetry breaking systematically, 
we need to vary the amount of symmetry beaking we have, while 
keeping all the flavour singlet terms in the action 
constant. We therefore follow a strategy in which the 
average quark mass $\mbar \equiv (m_u+m_d+m_s)/3$
is kept constant, while the mass splitting is increased.
Our notation for the quark masses and their splittings is 
\begin{eqnarray}
      \mbar &\equiv& \frac{1}{3} (  m_u + m_d + m_s)  \quad {\rm fixed }
                                                     \nonumber  \\
      \muh  &\equiv& m_u -\mbar                      \nonumber  \\
      \mdh  &\equiv& m_d -\mbar                                 \\
      \msh  &\equiv& m_s -\mbar                      \nonumber   \\
        m_l &\equiv &\frac{1}{2} ( m_u + m_d )        \nonumber  \\
        \mlh& \equiv& m_l -\mbar\,.                  \nonumber  
\label{dm_defs}
 \end{eqnarray}  
From these definitions we have the identity $\muh + \mdh + \msh = 0 $. 
In this notation the quark mass matrix is
\begin{eqnarray}
   {\cal M} &=&  \pmatrix{  m_u & 0 & 0 \cr 0 & m_d & 0 \cr 0 & 0 & m_s}
      \nonumber \\
            &=& \mbar\pmatrix{  1 & 0 & 0 \cr 0 & 1 & 0 \cr 0 & 0 & 1 }
                + {\ts \frac{1}{2} } (\muh - \mdh)
   \pmatrix{  1 & 0 & 0 \cr 0 & -1 & 0 \cr 0 & 0 & 0 }
                + {\ts \frac{1}{2}} \msh \pmatrix{ -1 & 0 & 0 \cr
                   0 &-1 & 0 \cr 0 & 0 & 2 } \,.
 \end{eqnarray}
$ {\cal M}$  has a flavour singlet part (proportional to $I$)
and a flavour octet part, proportional to $\lambda_3$, $\lambda_8$.
It is important to note that there are no terms in the QCD
Lagrangian which are in representations higher than the octet. 
The only way to give a value to a quantity in a higher $SU(3)$ 
representation is to have multiple powers of the flavour-breaking term,
i.e.\ multiple powers of $\delta m_q$. 
 
Therefore we adopt the following strategy. 
We classify physical quantities by their representation of $SU(3)$
and its sub-group $SU(2)$, and classify quark mass polynomials in 
the same way. The Taylor expansion of a quantity of known symmetry
can only involve polynomials of the matching symmetry. This 
strongly constrains the Taylor expansion of physical quantities 
about a symmetric point with all three quark masses equal. 

\begin{figure}[h]
   \begin{center}
      \includegraphics[width=5.50cm]{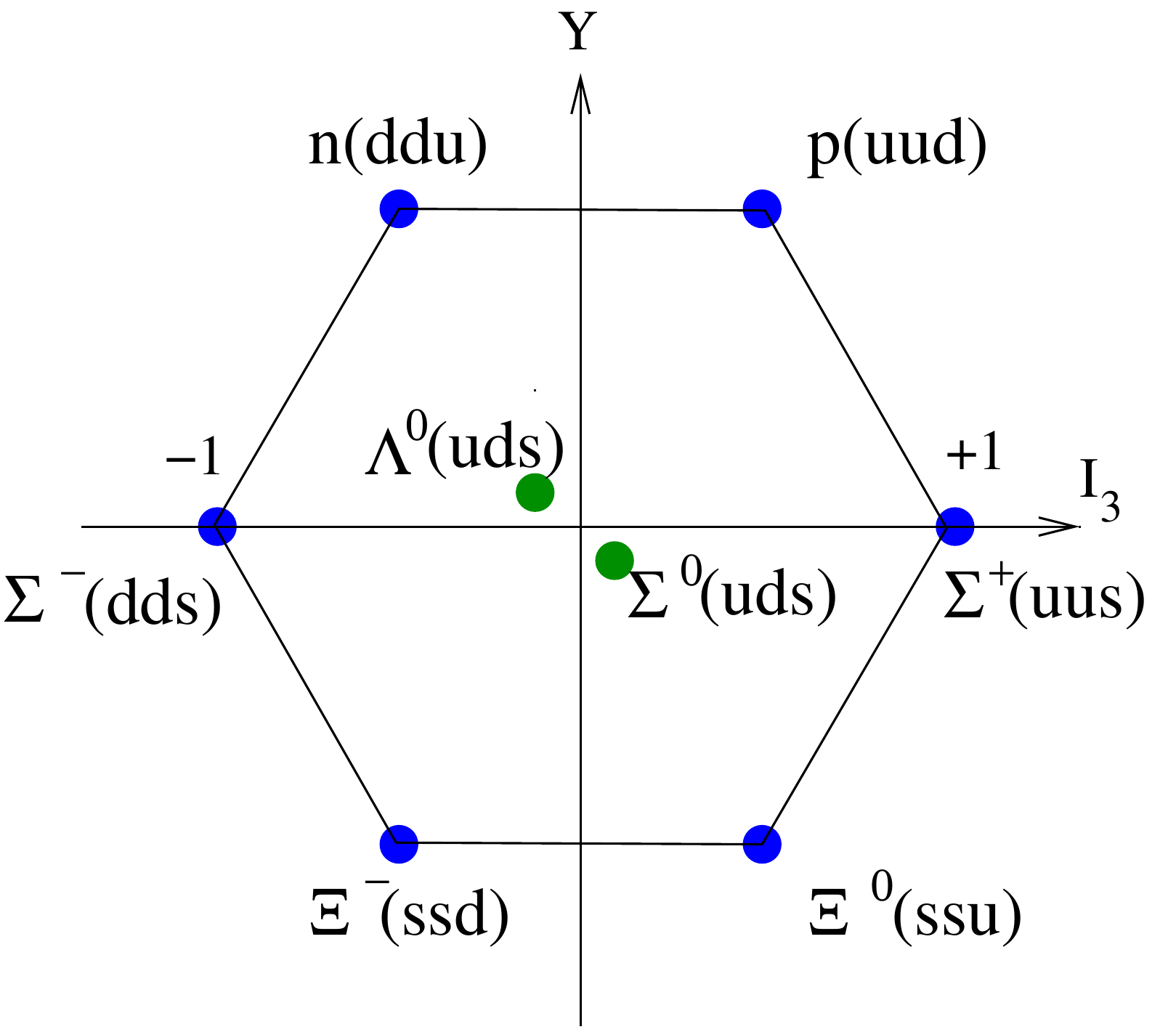}
   \end{center} 
\caption{The (lowest) octet baryon states.}
\label{octet_reps}
\end{figure}

In this work we are investigating non-singlet matrix elements
(e.g.\ vector and axial-vector currents for weak decays) acting
between octet baryons. This octet is illustrated in Fig.~\ref{octet_reps}.
The matrix elements have the form $\langle B^i | O^j | B^k \rangle$
with  $i,j,k \in \{1, \cdots, 8\} $ because the hadrons and operators
are both in flavour octets. Our choice of indices is set out in
Table~\ref{ind8}. We use the corresponding meson name to refer
\begin{table}[htb]
   \begin{center}
      \begin{tabular} {|c|ccc|} \hline
 Index & Baryon\,($B$) & Meson\,($M$) & Operator\,($O$) \cr \hline
 1     & $n$ & $K^0$ & $ \bar d \gamma s $ \tophat \\
 2     & $p$ & $K^+$ & $ \bar u \gamma s $  \\
 3     & $\Sigma^-$ &  $\pi^-$ & $ \bar d \gamma u $  \\
 4     & $\Sigma^0$ & $\pi^0$ & $
   \frac{1}{\sqrt{2}}\left( \bar u \gamma u - \bar d \gamma d \right) $ \\
 5     & $\Lambda^0$ &  $\eta$  & $\frac{1}{\sqrt{6}}\left(
  \bar u \gamma u +  \bar d \gamma d -2  \bar s \gamma s \right) $ \\
 6     & $\Sigma^+$ & $\pi^+$ & $\bar u \gamma d $ \\
 7    & $\Xi^-$ & $K^-$ & $\bar s \gamma u $ \\
 8    & $\Xi^0$ & $\bar K^0$ & $\bar s \gamma d $  \\ \hline
      \end{tabular}
\caption{Our numbering of the octet states, used internally.
         Whenever possible, final results will be presented in ways
         that are independent of the index choice. We use the convention
         that operator number $i$ has the same effect as absorbing a
         meson with the index $i$.}
\label{ind8}
\end{center}
\end{table}
to the flavour of bilinear quark operators, for example
\begin{eqnarray} 
   \langle p | \pi^0 | p \rangle
     \equiv \langle p | \frac{1}{\sqrt{2}} ( \overline{u} \gamma u -
                        \overline{d} \gamma d ) |p \rangle
     \equiv \langle B^2 | O^4 | B^2 \rangle \,,
\end{eqnarray} 
where $\gamma$ is a generic gamma matrix. Since we need three indices,
$i$, $j$, $k$, to specify a matrix element, we now need to classify
$8 \times 8 \times 8$ tensors under $SU(3)$, in just the same way
as we needed to give the classification of $8 \times 8$ and
$10 \times 10$ matrices for baryon masses.

To find the allowed mass-dependence of octet matrix elements of octet
hadrons we need the $SU(3)$ decomposition of $8 \otimes 8 \otimes 8$.
Using the intermediate result
\begin{eqnarray}
   8 \otimes 8 = 1 \oplus 8 \oplus 8 \oplus 10 \oplus  {\overline {10}}
                 \oplus 27 \,,
\label{decomp88}
\end{eqnarray}  
we find
\begin{eqnarray}
   \! \! 8 \otimes 8 \otimes 8 \!\! &=& \!\! 1 \oplus 1
    \oplus 8 \oplus 8 \oplus 8 \oplus 8 \oplus 8 \oplus 8 \oplus 8 \oplus 8
            \nonumber \\ && \! \!
    \oplus 27  \oplus 27  \oplus 27  \oplus 27  \oplus 27  \oplus 27
          \oplus{64}
            \nonumber \\ && \! \!
    \oplus 10 \oplus 10 \oplus 10 \oplus 10
          \oplus {\overline {10}}  \oplus {\overline {10}}
    \oplus {\overline {10}}  \oplus {\overline {10}}
            \nonumber \\ && \! \!
    \oplus 35  \oplus 35 \oplus{\overline{35}} \oplus{\overline{35}} \,.
\label{decomp888}
\end{eqnarray}

The allowed quark mass Taylor expansion for a hadronic matrix element
must follow the schematic pattern
\begin{eqnarray}
   \langle H^i | O^j | H^k \rangle &=&
      \sum (\mbox{singlet mass polynomial}) \times
            (\mbox{singlet tensor})^{ijk} \nonumber  \\
      && + \sum (\mbox{octet mass polynomial}) \times
            (\mbox{octet tensor})^{ijk} \label{schema} \\
      && + \sum (\mbox{27-plet mass polynomial}) \times
            (\mbox{27-plet tensor})^{ijk} \nonumber  \\
      && + \sum (\mbox{64-plet mass polynomial}) \times
            (\mbox{64-plet tensor})^{ijk} \nonumber  \\
      && + \ \cdots\,.  \nonumber
\end{eqnarray}
The tensors in this equation are three-dimensional arrays of
integers and square-roots of integers, objects somewhat analogous to
three-dimensional Gell-Mann matrices.

Mass polynomials with the symmetry $10$, $\overline{10}$, $35$,
$\overline{35}$ all have factors of $(m_u - m_d)$. So they only
appear if we consider the $1 + 1 + 1$ case of symmetry breaking.
At present we are only considering the $2+1$ case, $m_u = m_d \ne m_s $
so we can neglect the $10$, $\overline{10}$, $35$, $\overline{35}$
representations.

We found just two singlet tensors in the expansion of
$8 \otimes 8 \otimes 8$, so at the symmetric point there
are only two independent coefficients (usually called
$F, D$ or $f, d$) needed to completely specify all the matrix
elements between the members of the octet.
These give the classic $SU(3)$ inter-relations between
octet amplitudes. These are generally found to work rather
well. We should however be able to do better by also
including higher terms in the mass expansion.

There are $8$ octets in the expansion of $8 \otimes 8 \otimes 8$,
so if we work to first order $\delta m_q$, the $SU(3)$ flavour violation,
we have $8$ new coefficients. There are still many fewer coefficients
than there are amplitudes, so there are numerous constraints
and cross-relations between amplitudes. The singlet and octet tensors
are given explicitly in Table~\ref{coef_1_8}. 
\begin{table}[htb]
   \begin{center}
      \begin{tabular}{|cc|rr|rrrrrrrr|} \hline
  & & \multicolumn{2}{c|}{1} & \multicolumn{8}{c|}{8} \\
 $I$ & $A_{\overline{B}^\prime M B}$ & $f$ & $d$ &$r_1$ &$r_2$ &$r_3$ &$r_4$ &$r_5$
 & $s_1$ & $s_2$ & $s_3$ \\
 \hline \hline
 $0$ &  \tophat ${\overline{N}\eta N}$ & $\sqrt{3}$ & $-1$
 & 1  & 0 & 0 & 0& 0& 0& $-1$ & 0 \\
 $0$ & ${\overline{\Sigma}\eta \Sigma}$ & 0& 2
 & 1 & 0 &  $2 \sqrt{3}$ & 0 & 0 & 0 & 0 & 0 \\
 $0$ & ${\overline{\Lambda}\eta \Lambda}$ & 0& $-2$
 & 1 & 2 & 0 & 0 & 0 & 0 & 0 & 0 \\
 $0$ & ${\overline{\Xi}\eta \Xi}$ &$-\sqrt{3} $ & $-1$
 & 1 & 0  & 0 & 0& 0& 0 & 1 & 0\\
 \hline
 $1$ & \tophat ${\overline{N}\pi N}$ & 1 & $\sqrt{3}$
 & 0& 0 & $-2$ & 0 & 0 & 2 & 0 & 0 \\
 $1$ &  ${\overline{\Sigma}\pi \Sigma}$ & 2 & 0
 & 0 & 0 & 0 & 0 & 0 &$-2$ &$\sqrt{3}$ & 0\\
 $1$ &  ${\overline{\Xi}\pi \Xi}$ & 1 & $-\sqrt{3}$
 & 0 & 0 & $2$ & 0 & 0 & 2 & 0 & 0 \\
 \hline \hline
 $1$ &  ${\overline{\Sigma}\pi \Lambda}$ & 0 & \tophat 2
 & 0 & 1 & $-\sqrt{3}$ & $i$ & 0 & 0 & 0 & 0 \\
  $1$ &  ${\overline{\Lambda}\pi \Sigma}$ & 0 & 2
  & 0 & 1 & $-\sqrt{3}$ & $-i$ & 0 & 0 & 0 & 0 \\
 \hline
 $\frac{1}{2}$ & ${\overline{N} K \Sigma}$ & $-\sqrt{2}$ & $\sqrt{6}$&
 \tophat
   0 & 0& $ \sqrt{2}$ & 0& $i\sqrt{2}$ & $\sqrt{2}$ & 0& $i\sqrt{6}$\\
 $\frac{1}{2}$ & ${\overline{N} K \Lambda}$ & $-\sqrt{3}$ & $-1$&
   0 & 1& 0 & $i$& $i\sqrt{3}$ & $-\sqrt{3}$ & 1& $-i$\\
 $\frac{1}{2}$ & ${\overline{\Lambda} K \Xi}$ & $ \sqrt{3}$ & $-1$
 & 0 & 1& 0 & $-i$& $-i\sqrt{3}$ & $\sqrt{3}$ & $-1$& $-i$\\
 $\frac{1}{2}$ & ${\overline{\Sigma} K \Xi}$ & $ \sqrt{2}$ & $\sqrt{6}$
 & 0 & 0& $ \sqrt{2}$ & 0& $-i\sqrt{2}$ & $-\sqrt{2}$ & 0& $i\sqrt{6}$\\
  \hline
  $\frac{1}{2}$ & ${\overline{\Sigma} \overline{K} N}$ & $-\sqrt{2}$ 
                                                       & $\sqrt{6}$&
  \tophat
    0 & 0& $ \sqrt{2}$ & 0& $-i\sqrt{2}$ & $\sqrt{2}$ & 0& $-i\sqrt{6}$\\
  $\frac{1}{2}$ & ${\overline{\Lambda} \overline{K} N}$ & $-\sqrt{3}$ & $-1$&
    0 & 1& 0 & $-i$& $-i\sqrt{3}$ & $-\sqrt{3}$ & 1& $i$\\
  $\frac{1}{2}$ & ${\overline{\Xi} \overline{K} \Lambda}$ & $ \sqrt{3}$ & $-1$
  & 0 & 1& 0 & $i$& $i\sqrt{3}$ & $\sqrt{3}$ & $-1$& $i$\\
  $\frac{1}{2}$ & ${\overline{\Xi} \overline{K} \Sigma}$ & $ \sqrt{2}$ 
                                                         & $\sqrt{6}$
  & 0 & 0& $ \sqrt{2}$ & 0& $i\sqrt{2}$ & $-\sqrt{2}$ & 0& $-i\sqrt{6}$\\
\hline \hline
\end{tabular}
\caption{Coefficients in the mass Taylor expansion of operator amplitudes:
         $SU(3)$ singlet and octet. These coefficients are sufficient
         for the linear expansion of hadronic amplitudes.}
\label{coef_1_8}
\end{center}
\end{table}
This table gives the amplitudes for the baryons $p$, $\Lambda^0$,
$\Sigma^+$, $\Xi^0$; the amplitudes for the other baryons can be deduced
from isospin symmetry (which we are, for now, treating as unbroken). 
We have used the notation for the matrix element transition
$B \to B^\prime$ of
\begin{eqnarray}
   A_{\overline{B}^\prime M B} = \langle B^\prime | M | B \rangle \,,
\end{eqnarray}
where $M$ is the appropriate operator from Table~\ref{ind8}.
We illustrate how this table works by reading off the mass expansion
for the first two amplitudes 
\begin{eqnarray}
   \langle p | \eta | p \rangle  = A_{\overline{N}\eta N} 
       &=& \sqrt{3} f - d + (r_1  - s_2) \dml                \nonumber \\
   \langle \Sigma^+ | \eta | \Sigma^+ \rangle = A_{\overline{\Sigma}\eta \Sigma}
       &=& 2 d  + (r_1 + 2 \sqrt{3} r_3 ) \dml               \nonumber \\
       &\vdots&
\label{matrix_eg}
\end{eqnarray} 
 

\section{`Fan' Plots} 


In the case of hadron masses we found that `fan' plots were a useful
way to display our results, see e.g.\ Fig.~20 of \cite{bietenholz11a}.
We plotted the masses of the hadrons in  a multiplet against
$\dml = m_l - \mbar$. When $\dml = 0$ (the $SU(3)$ symmetric point)
all masses are equal, as we increase the symmetry breaking
the masses fan out, about an `average' mass which is almost constant. 
 
We can display matrix elements in a similar way. Some matrix 
elements, however, will be protected from first order flavour symmetry 
breaking effects. The Ademollo--Gatto theorem, \cite{ademollo64a},
for example, states that certain form factors for vector currents
will not display $\dml$ effects.


\subsection{The $f$-fan}


Using Table~\ref{coef_1_8}  we can construct five quantities $F_i$,
which all have the same value ($2f$) at the symmetric point, but which
can differ once $SU(3)$ is broken.
\begin{eqnarray}
   F_1 \equiv \frac{1}{\sqrt{3}} ( A_{\bar N \eta N} - A_{\bar \Xi \eta \Xi} )
       &=& 2 f - \frac{2}{\sqrt{3}} s_2 \dml \nonumber \\
   F_2 \equiv  ( A_{\bar N \pi  N} + A_{\bar \Xi \pi  \Xi} )
       &=& 2 f + 4 s_1 \dml \nonumber \\
   F_3 \equiv A_{\bar \Sigma \pi \Sigma }
       &=& 2 f + (-2 s_1 + \sqrt{3} s_2) \dml \\ 
   F_4 \equiv \frac{1}{\sqrt{2}} \Re ( A_{\bar \Sigma K \Xi} -A_{\bar N K \Sigma} )
       &=& 2 f - 2 s_1 \dml \nonumber \\
   F_5 \equiv \frac{1}{\sqrt{3}} \Re ( A_{\bar \Lambda K \Xi}
                                       -A_{\bar N K \Lambda})
       &=& 2 f + \frac{2}{\sqrt{3}} (\sqrt{3} s_1 - s_2) \dml \,. \nonumber
\label{F_expansion}
\end{eqnarray}   
Plotting these quantities gives a `fan' plot with $5$ lines, but only $2$
slope parameters ($s_1$, $s_2$), so the splittings between these observables
are highly constrained.
 
A useful `average F' can be constructed from the diagonal amplitudes
\begin{eqnarray}
   X_F = \frac{1}{6}(3F_{1} + F_{2} + 2F_{3}) = 2f + O(\dml^{2}) \,.
\label{XF}
\end{eqnarray}
We expect that a `fan' plot of $\tilde{F}_i \equiv F_{i}/X_F$ might be less
noisy than a plot of $F_{i}$ alone. (Using $\tilde{F}_i$ rather than
$F_i$ would also remove renormalisation constants, see eq.~(\ref{tilde_rats}).)
In general we shall denote quantities with a tilde that have been
normalised with an appropriate $X$.


\subsection{The $d$-fan}


Similarly, we can construct seven quantities $D_i$, which all have
the same value ($2d$) at the symmetric point, but which can differ
once $SU(3)$ is broken.
\begin{eqnarray}
   D_1 \equiv - ( A_{\bar N \eta N} + A_{\bar \Xi \eta \Xi} )
     &=& 2 d - 2 r_1 \dml \nonumber \\
   D_2 \equiv A_{\bar \Sigma \eta \Sigma }
     &=& 2 d + (r_1 + 2 \sqrt{3} r_3) \dml \nonumber \\
   D_3 \equiv{} - A_{\bar \Lambda \eta \Lambda }
     &=& 2 d - (r_1 + 2 r_2) \dml \nonumber \\
   D_4 \equiv \frac{1}{\sqrt{3}} ( A_{\bar N \pi  N} - A_{\bar \Xi \pi  \Xi} )
     &=& 2 d -\frac{4}{\sqrt{3}} r_3 \dml \\ 
   D_5 \equiv \Re A_{\bar \Sigma \pi \Lambda }
     &=& 2 d + ( r_2 - \sqrt{3} r_3) \dml \nonumber \\
   D_6 \equiv \frac{1}{\sqrt{6}} \Re (A_{\bar N K \Sigma} + A_{\bar \Sigma K \Xi} )
     &=& 2 d + \frac{2}{\sqrt{3}} r_3 \dml \nonumber \\
   D_7 \equiv - \Re ( A_{\bar N K \Lambda} + A_{\bar \Lambda K \Xi} )
     &=& 2 d - 2 r_2 \dml\,. \nonumber
\label{D_expansion}
\end{eqnarray}
Plotting these quantities gives a `fan' plot with $7$ lines, but only $3$
slope parameters ($r_1, r_2$ and $r_3$), so once again the splittings
between these observables are highly constrained. Again it is possible
to construct an `average D' similar to $X_F$ 
\begin{eqnarray}
   X_D = \frac{1}{4}(D_{1} + 2D_{2} + D_{4}) = 2d + O(\dml^{2}) \,,
\label{XD}
\end{eqnarray}
in order to produce a less noisy `fan' plot.


\section{Lattice Calculations} 


In order to extract the matrix elements for some operator $O$,
it is necessary to take an appropriate ratio of three and two-point
correlation functions, \cite{martinelli89a,wilcox92a}
\begin{eqnarray}
   {\cal R} 
      = \frac{C_{3}^{B \rightarrow B^\prime}(t,\tau; p,p^\prime)}
                                             {C_{2}^{B^\prime}(t,p^\prime)}
           \sqrt{\frac{C_{2}^{B^\prime}(t,p^\prime)C_{2}^{B^\prime}(\tau,p^\prime)
                                C_{2}^{B}(t-\tau,p)}
                {C_{2}^{B}(t,p)C_{2}^{B}(\tau,p)
                                C_{2}^{B^\prime}(t-\tau,p^\prime)}} \,,
\end{eqnarray}
where
\begin{eqnarray}
   C_{3}^{B\to B^\prime}(t,\tau; p, p^\prime) 
      &=& \mbox{tr}_D\,\Gamma
            \langle {\cal B}_{B^\prime}(t;\vec{p}^\prime) O(\tau;\vec{q})
                                 \overline{\cal B}_B(0;\vec{p}) \rangle 
                                                         \nonumber   \\
   C_{2}^B(t,p) 
      &=& \mbox{tr}_D\,\Gamma_{unpol}
            \langle {\cal B}_B(t;\vec{p}) \overline{\cal B}_B(0;\vec{p})
                    \rangle \,.
\end{eqnarray}
This is designed so that any smearing for the source (at time $0$)
and sink operators (at time $t$) is cancelled in the ratios,
\cite{capitani98a}; of course smearing improves the overlap with
the lowest lying state. The baryon operators used are as follows
\begin{eqnarray}
   {\cal B}_{N\,\alpha}(t;\vec{p})
      &=& \sum_{\vec{x}} 
             e^{-i\vec{p}\cdot\vec{x}}\epsilon_{ijk} u_{i \alpha}(\vec{x},t)
                         (u^{T_{D}}_{j}(\vec{x},t) C \gamma_{5} d_{k}(\vec{x},t))
                                                         \nonumber   \\
   {\cal B}_{\Sigma\,\alpha}(t;\vec{p})
      &=& \sum_{\vec{x}}
             e^{-i\vec{p}\cdot\vec{x}}\epsilon_{ijk} u_{i \alpha}(\vec{x},t)
                         (u^{T_{D}}_{j}(\vec{x},t) C \gamma_{5} s_{k}(\vec{x},t))
                                                         \nonumber   \\
   {\cal B}_{\Lambda\,\alpha}(t;\vec{p})
      &=& \sum_{\vec{x}}
             e^{-i\vec{p}\cdot\vec{x}}\epsilon_{ijk} s_{i \alpha}(\vec{x},t)
                          (u^{T_{D}}_{j}(\vec{x},t) C \gamma_{5} d_{k}(\vec{x},t))
                                                         \nonumber   \\
   {\cal B}_{\Xi\,\alpha}(t;\vec{p}) 
      &=& \sum_{\vec{x}}
             e^{-i\vec{p}\cdot\vec{x}}\epsilon_{ijk} s_{i \alpha}(\vec{x},t)
                   (s^{T_{D}}_{j}(\vec{x},t) C \gamma_{5} u_{k}(\vec{x},t)) \,,
\label{eq:interpolating}
\end{eqnarray}
where $C$ is the charge conjugation matrix and $i,j,k$ are colour
indices and $\alpha$ is a Dirac index. The $u$ and $d$ quarks are
treated distinctly, but with degenerate mass. The transferred momentum
from the initial, $B$, to final, $B^\prime$ state is given by
\begin{eqnarray}
  q = p - p^\prime = \left( i(E_B(\vec{p}) - E_{B^\prime}(\vec{p}^\prime)),
                          \vec{p} - \vec{p}^\prime \right) \,.
\end{eqnarray}
In this study we shall restrict ourselves to zero $3$-momentum transfer,
i.e.\ $\vec{p} = \vec{p}^\prime = 0$ when
\begin{eqnarray}
   q^2 \to q^{2}_{max} = -(M_B-M_{B^\prime})^{2} \,,
\end{eqnarray}
such that postive $q^{2}$ is spacelike while negative $q^{2}$ is a timelike quantity. 
The energy of the initial and final states are now simply
the rest masses $M_B$ and $M_{B^\prime}$ respectively.
We shall also consider the time-like component of the vector
and axial-vector currents as described in Table~\ref{ind8} by taking
$M = \pi^0$ (or $\pi^+$) and $\gamma = \gamma_4$ or $M = K^+$ and
$\gamma = \gamma_3 \gamma_{5}$ respectively. For example we can set $O$ to be 
\begin{eqnarray}
   \left. \begin{array}{ccc}
             V_4 &=& \overline{u}\gamma_4 d \\
             A_3 &=& \overline{u}\gamma_3\gamma_5 d \\
          \end{array} \right\} \,\, \mbox{for $\Delta S = 0$ decays}\,,
   \qquad
   \left. \begin{array}{ccc}
             V_4 &=& \overline{u}\gamma_4 s \\
             A_3 &=& \overline{u}\gamma_3\gamma_5 s \\
          \end{array} \right\} \,\, \mbox{for $\Delta S = 1$ decays}\,.
\end{eqnarray}
Even though in this case $\vec{p} = \vec{p}^\prime = 0$, in 
general there is still a $4$-momentum transfer and we usually have a 
non-forward matrix element. Thus we have
\begin{eqnarray}
   {\cal R} \rightarrow A_{\overline{B}^\prime M B}(q^{2}_{max}) \,,
     \quad \mbox{as} \quad t\,,\, t-\tau \to \infty\,,
\end{eqnarray}
depending on whether we are considering the vector $O = V_4$,
$\Gamma = \Gamma_{unpol}$ or axial-vector $O = A_3$, $\Gamma = \Gamma_{pol}$
three-point function.

The computed matrix elements are bare (or lattice) quantities
and must be renormalised,
\begin{eqnarray}
   V_4^{\R} = Z_V V_4 \,, \qquad A_3^{\R} = Z_A A_3 \,,
\end{eqnarray}
where we have denoted the renormalised matrix elements with a
superscript $\mbox{R}$.
If $A^{\R}_{\overline{B}^\prime M^V B}$ is known then the renormalisation
constant can simply be determined from
\begin{eqnarray}
   Z_V = { A^{\R}_{\overline{B}^\prime M^V B} \over A_{\overline{B}^\prime M^V B} } \,, \qquad
   Z_A = { A^{\R}_{\overline{B}^\prime M^A B} \over A_{\overline{B}^\prime M^A B} } \,.
\label{renorm_det}
\end{eqnarray}
Alternatively by considering ratios in the `fan' plots the renormalisation
constant cancels, for example
\begin{eqnarray}
   \tilde{F}^A_i = {F_i^A \over X_{F^A}} 
                 = {F_i^{A\,\R} \over X^{\R}_{F^A}} = \tilde{F}_i^{A\,\R}\,,
         \quad i = 1\, \ldots\,, 5 \,,
\label{tilde_rats}
\end{eqnarray}
and similarly for $\tilde{F}_i^V$, $\tilde{D}_i^A$ and $\tilde{D}_i^V$.

The renormalised vector $f$ and $d$ coefficients
are known (i.e.\ the matrix elements at the $SU(3)$ flavour symmetric
point). The vector current, being conserved there, essentially just
counts the number of quarks. For example using eq.~(\ref{matrix_eg})
we have
\begin{eqnarray}
   A_{\overline{N}\eta^V N}^{\R} 
      &=& {1\over \sqrt{6}}(2+1-0) = \sqrt{3}f_V^{\R} - d_V^{\R} \,,
                                                     \nonumber  \\
   A_{\overline{\Sigma}\eta^V \Sigma}^{\R} 
      &=& {1\over \sqrt{6}}(2+0-2) = 2d_V^{\R} \,,
\end{eqnarray}
giving
\begin{eqnarray}
   f_V^{\R} = {1 \over \sqrt{2}}\,, 
   \qquad \mbox{and}\quad 
   d_V^{\R} = 0 \,.
\label{fVR}
\end{eqnarray}
This result can be used to estimate the renormalisation constant.
Either eq.~(\ref{renorm_det}) can be used at the symmetric
point or equivalently if we have measured $X_{F_V}$ 
($\gamma = \gamma_4$) then from eq.~(\ref{XF})
we have up to $O(\delta m_l^2)$
\begin{eqnarray}
   Z_V = {X_{F_V}^{\R} \over X_{F_V}}
       = {2f_V^{\R} \over X_{F_V}} = {\sqrt{2} \over X_{F_V}} \,,
\label{ZV_determin}
\end{eqnarray}
For the axial current we can connect our conventions with others
in the literature, e.g.\ \cite{gaillard84a,mateu05a}, via
\begin{eqnarray}
   f_A^{\R} = {1\over \sqrt{2}}F_A \,,
   \qquad \mbox{and}\quad 
   d_A^{\R} = {1\over \sqrt{6}}D_A \,,
\end{eqnarray}
at the symmetric point. Similarly to $Z_V$ for $Z_A$ again either
eq.~(\ref{renorm_det}) can be used, for example for neutron $\beta$-decay
\begin{eqnarray}
   A_{\overline{N}KN} = {1 \over \sqrt{2}} g_{A}^{\R} \,,
\end{eqnarray}
where $g_{A}^{\R}$ the axial-vector coupling in
$\beta$-decay at the physical point. Equivalently from measuring
$X_{F_A}$ and $X_{D_A}$ yields
\begin{eqnarray}
   Z_A = {X_{F_A}^{\R} \over X_{F_A}}
         = {2f_A^{\R} \over X_{F_A}} 
         \,,\qquad \mbox{or} \quad
   Z_A = {X_{D_A}^{\R} \over X_{D_A}}
         = {2d_A^{\R} \over X_{D_A}} \,,
\label{ZA_determin}
\end{eqnarray}
(provided that either $X_{F_A}^{\R}$ or $X_{D_A}^{\R}$ is known).
Alternatively the ratio where $Z_A$ cancels is
\begin{eqnarray}
   {f_A \over d_A} =  { X_{F_A} \over X_{D_A} } \,.
\label{rat_fod}
\end{eqnarray}


\section{Results}


In a similar manner to previous simulations \cite{bietenholz10a,bietenholz11a}
our gauge field configurations have been generated with $N_f = 2+1$
flavours of dynamical fermions, using the tree-level Symanzik improved
gluon action and nonperturbatively $O(a)$ improved Wilson fermions,
\cite{cundy09a}. The quark masses are chosen by first finding the
$SU(3)$ flavour symmetric point where flavour singlet quantities
take on their physical values and vary the individual quark masses
while keeping the singlet quark mass
$\overline{m} = (m_u+m_d+m_s)/3 = (2m_l+m_s)/3$ constant, as described
in section~\ref{su3_breaking}. Simulations are performed on lattice
volumes of $24^{3} \times 48$ at $\beta = 5.50$ corresponding to a
lattice spacing of $a \approx 0.079$, \cite{bietenholz11a}.
Our calculations are performed on $5$ ensembles chosen by methods
as also outlined in \cite{bietenholz11a} presently using
$\sim 400$--$500$ trajectories for off--diagonal matrix elements
$B^\prime \not= B$ and $\sim 2000$ trajectories for diagonal matrix
elements $B^\prime = B$. In Table~\ref{ensembles}
\begin{table}[htb]
\begin{center}
   \begin{tabular}{|c|c|c|}
      \hline
      Ensemble & $\kappa_{l}$ & $\kappa_{s}$  \\
      \hline
      1 & 0.12083 & 0.12104  \\
      2 & 0.12090 & 0.12090  \\
      3 & 0.12095 & 0.12080 \\
      4 & 0.12100 & 0.12070 \\
      5 & 0.12104 & 0.12062 \\
      \hline
   \end{tabular}
\caption{Ensembles used in calculations here.}
\label{ensembles}
\end{center}
\end{table}
we give these $(\kappa_l, \kappa_s)$ values used here. Ensemble $2$
corresponds to the symmetric point. Note that for ensemble $1$
we have a universe where the $l$ quarks are heavier than the $s$ quark.

In \cite{horsley12a} (see also \cite{bietenholz11a}) the value of
the distance away from the symmetric point to the physical point
was determined to be $\delta m_l^* = -0.01102(3)$ in lattice units.
(The $*$ denotes the physical point; see eq.~(\ref{dm_defs})
for the definition of $\delta m_l$.)


\subsection{$X_D$, $X_F$}

\begin{figure}[h]
   \begin{center}
      \includegraphics[width=9.00cm]{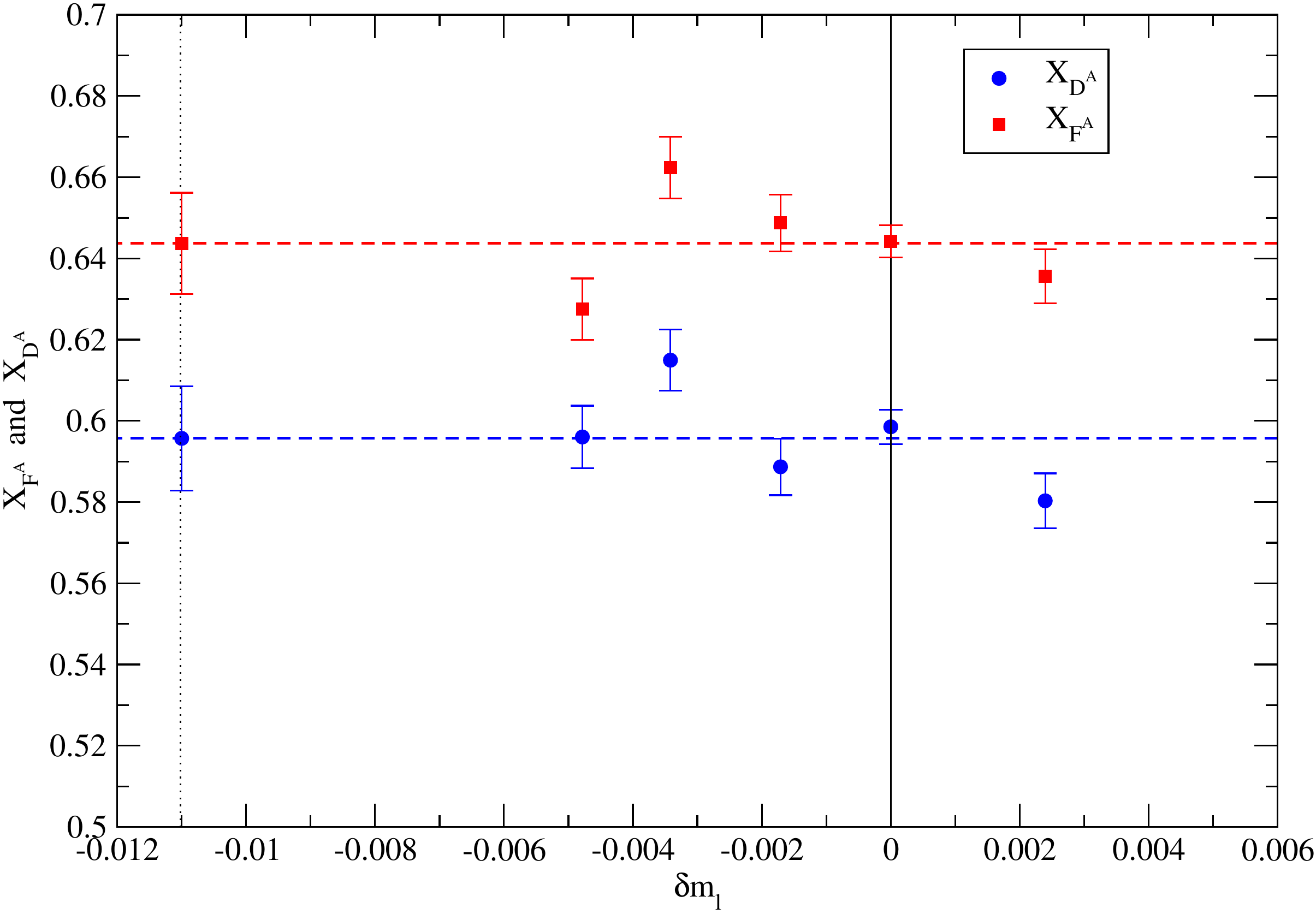}
   \end{center} 
\caption{$X_{F^A}$ (filled squares), $X_{D^A}$ (filled circles)
         against $\delta m_l$ each together with a constant fit.}
\label{X_plot}
\end{figure}

We first consider $X_F$, eq.~(\ref{XF}), and $X_D$, eq.~(\ref{XD})
for the axial-vector case. In Fig.~\ref{X_plot}
we plot $X_{F^A}$, $X_{D^A}$ against $\delta m_l$. We expect these
quantities to be constant (up to $O(\delta m_l^2)$ terms) and within
error bars this is indeed the case. From the values of the constant
fits and using eq.~(\ref{rat_fod}), we find
$f_A / d_A = 1.08(3)$. As it is well known that axial current matrix
elements suffer from large finite size effects, we are presently
repeating the determination of the matrix elements on larger
$32^3\times 64$ lattices.


\subsection{`Fan' plots}

\begin{figure}[htb]
\begin{minipage}{0.40\textwidth}

      \begin{center}
         \includegraphics[width=7.50cm]{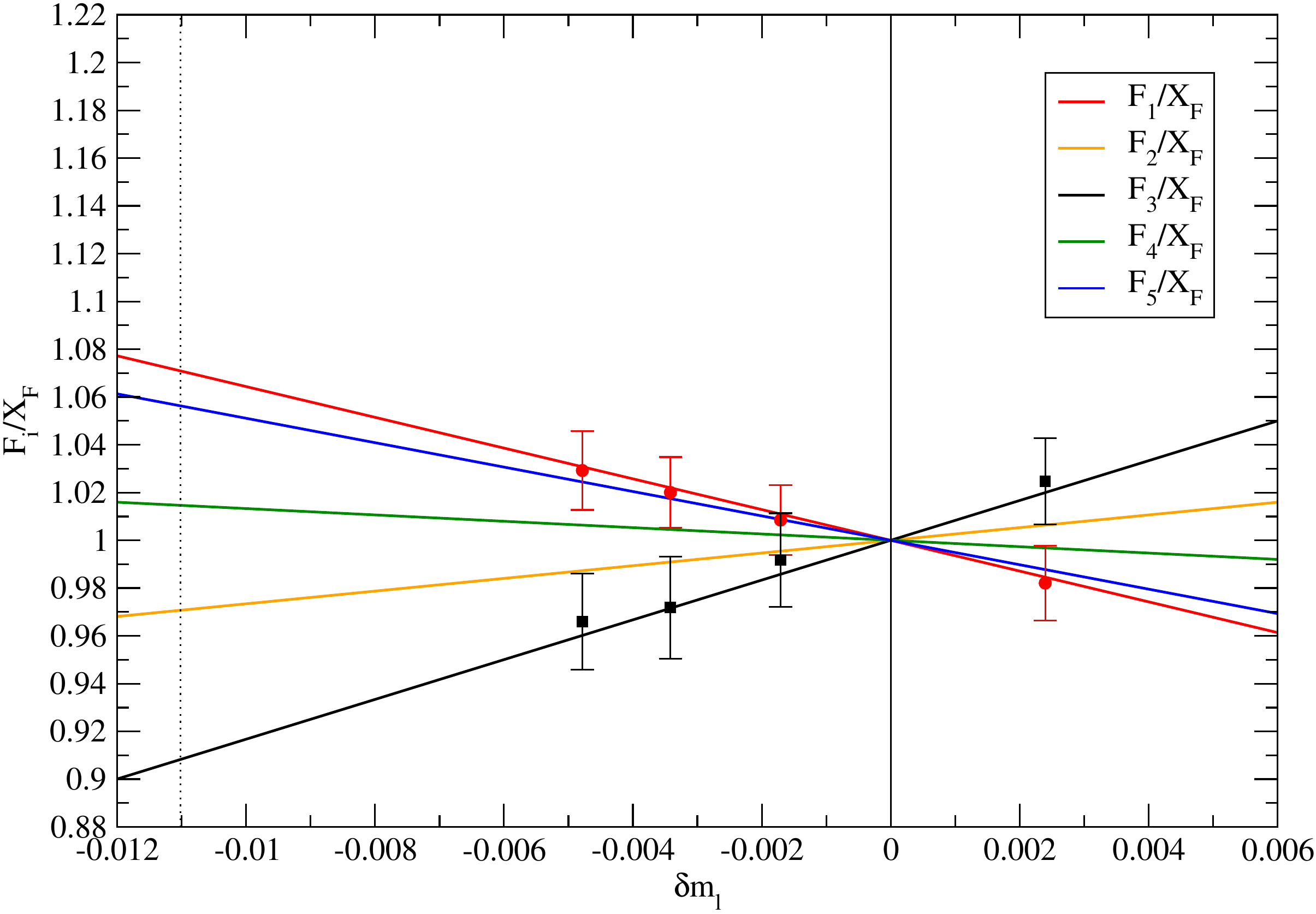}
      \end{center} 

\end{minipage} \hspace*{0.10\textwidth}
\begin{minipage}{0.40\textwidth}

      \begin{center}
         \includegraphics[width=7.50cm]{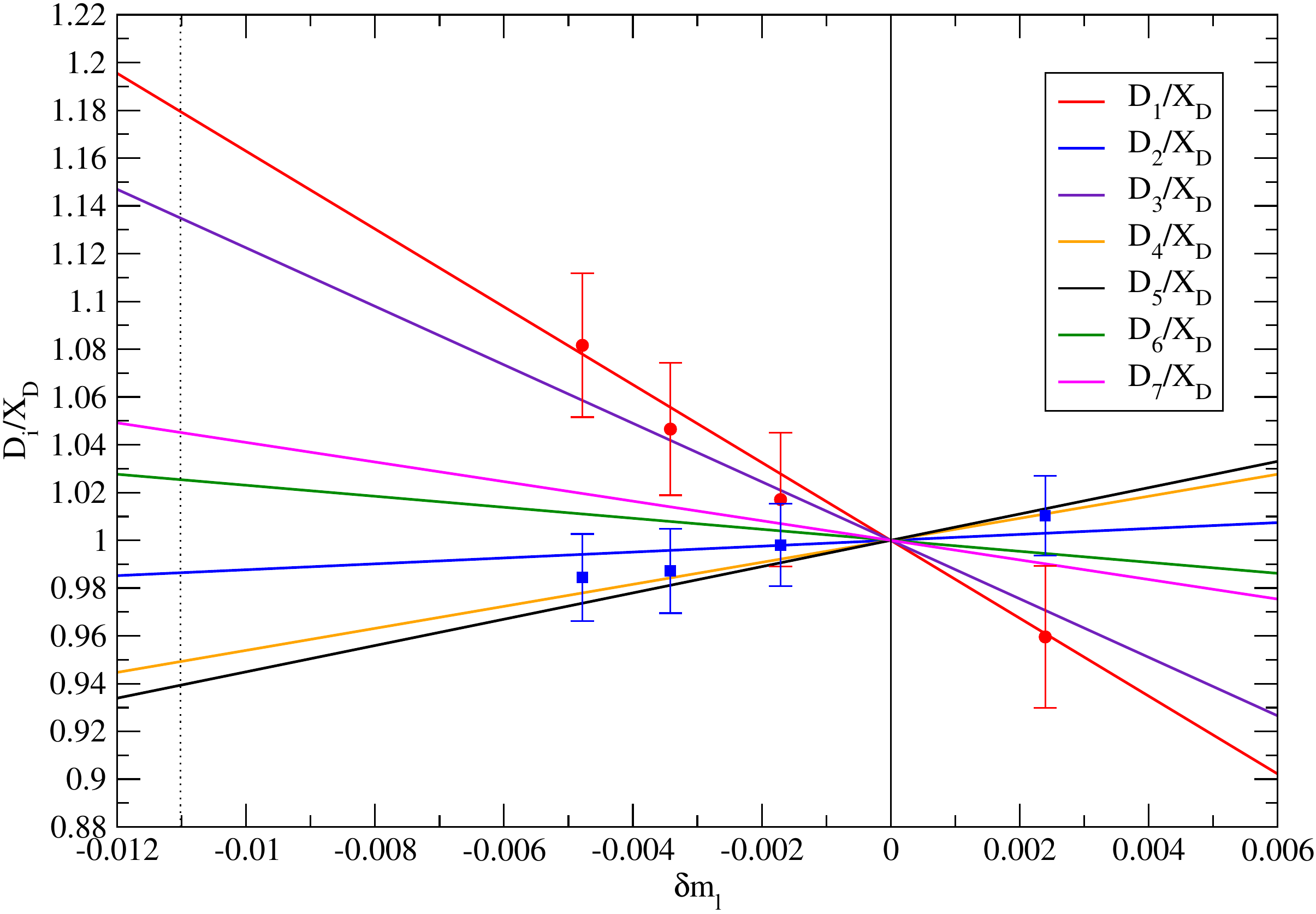}
      \end{center} 

\end{minipage}

\caption{Left panel: $\tilde{F}^A_i = F^A_i/X_{F_A}$ ($i = 1,\,\ldots,\,5$),
         right panel: $\tilde{D}^A_i = D^A_i/X_{D_A}$ ($i = 1,\,\ldots,\,7$).
         Shown are just the numerical results for $\tilde{F}^A_1$ (circles),
         $\tilde{F}^A_2$ (squares) and $\tilde{D}^A_1$ (circles),
         $\tilde{D}^A_3$ (squares) against $\delta m_l$ together with
         the (normalised) fits from
         eqs.~(\protect\ref{D_expansion}), (\protect\ref{F_expansion}).}

\label{AF_AD_plot}
\end{figure}

We now turn to a discussion of `fan' plots. As noted previously
\cite{bietenholz11a}, it is better to consider ratios, which are
less noisy. In Fig.~\ref{AF_AD_plot} we show $\tilde{F}^A_i = F^A_i/X_{F_A}$ 
(left panel) and $\tilde{D}^A_i = D^A_i/X_{D_A}$ (right panel)
together with representive numerical results. (Due to our relatively
low number of configurations used in the analysis, error bars
overlap if all numerical results are plotted.) Note that because
we have normalised the data with $X_{F^A}$ or $X_{D^A}$ so at the
symmetric point the ratios are $1$ exactly.

We make a simultaneous fit to the data to arrive at a determination for the
constants $\tilde{s}^A_1$, $\tilde{s}^A_2$, $\tilde{s}^A_3$ and $\tilde{r}^A_1$,
$\tilde{r}^A_2$, (where $\tilde{s}^A_i = s^A_i/2f^A$,
$\tilde{r}^A_i = r^A_i/2d^A$) based upon eqs.~(\ref{F_expansion}),
(\ref{D_expansion}).
Using these parameters it is then possible to reconstruct the equations
describing the $SU(3)$ flavour symmetry breaking effects on matrix elements
up to $O(\delta m_l)$. It should be noted that we have yet to perform
calculations for certain correlators, though we are progressing in this
direction \cite{zanottilat2012}. Those missing include 
$A_{\bar{\Sigma}K^{A}\Xi}$ which then precludes the use of
$F^A_4$ and $D^A_6$ in our simultaneous fits. The axial-vector matrix
elements clearly have linear terms in $\delta m_l$.

\begin{figure}[htb]
   \begin{center}
      \includegraphics[width=10.00cm]{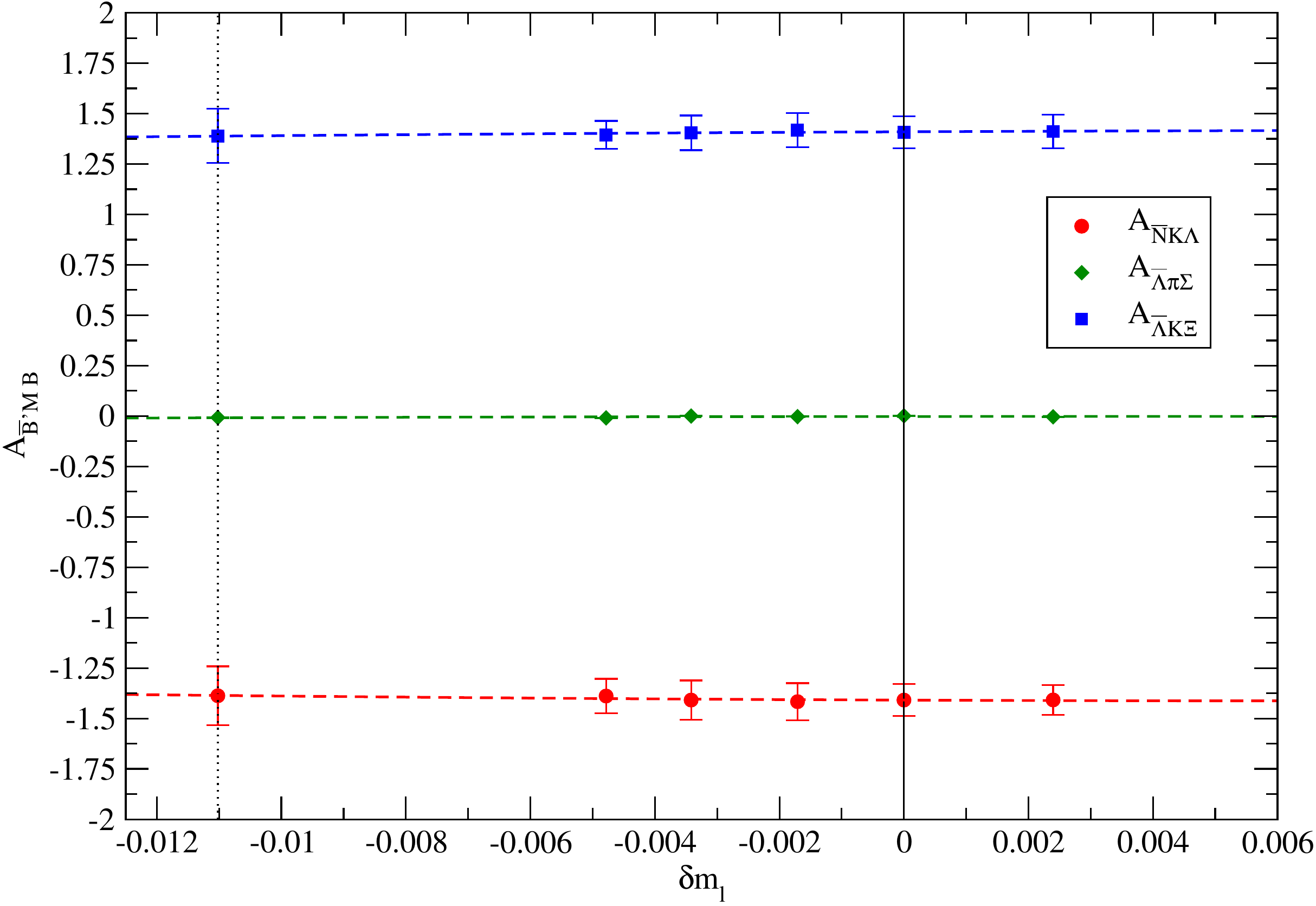}
   \end{center} 
\caption{The matrix elements $A_{\overline{N}K^V\Lambda}$
         (red filled circles), $A_{\overline{\Lambda}\pi^V\Sigma}$
         (green filled diamonds) and $A_{\overline{\Lambda}K^V\Xi}$
         (blue filled squares) against $\delta m_l$, together with
         a quadratic fit.}
\label{fig_f0}
\end{figure}

For the vector case due to the Ademello--Gatto theorem,
\cite{ademollo64a}, the linear terms in $\delta m_l$ are absent.
So instead of a `fan' plot we show in Fig.~\ref{fig_f0} a 
selection of matrix elements
$A_{\overline{N}K^V\Lambda}$, $A_{\overline{\Lambda}\pi^V\Sigma}$, and
$A_{\overline{\Lambda}K^V\Xi}$. As a check, from Table~\ref{coef_1_8}
we see that to leading order
$A^{\R}_{\overline{\Lambda}\pi^V\Sigma} = 0 + d^{\R}_V$ i.e.\ it
contains only a $d^{\R}_V$ term and no $f^{\R}_V$ term. However $d_V^{\R} = 0$,
eq.~(\ref{fVR}), so we expect to $O(\delta m_l^2)$ that
$A_{\overline{\Lambda}\pi^V\Sigma}$ vanishes, which is clearly seen
in Fig.~\ref{fig_f0}. The other decays are also flat. At the
symmetric point (or indeed at other points) we can estimate $Z_V$ from
eq.~(\ref{renorm_det}). As 
$A^{\R}_{\overline{N}K^V\Lambda} = - \sqrt{3/2} = - A^{\R}_{\overline{\Lambda}K^V\Xi}$
then we expect one result in Fig.~(\ref{fig_f0}) to be the mirror
image of the other. This is the case and choosing $A_{\overline{N}K^V\Lambda}$
gives upon using eq.~(\ref{renorm_det}), $Z_V = 0.87(5)$.


\section{Octet hyperon semi-leptonic decays}


The theory outlined in previous sections is general; most phenomenological
calculations are directed towards the semi-leptonic decays
$B \to B^\prime e\overline{\nu}_e$ of various octet hyperons in order to help
determine $|V_{us}|$, e.g.\ \cite{cabibbo03a}. We now briefly indicate
how far our programme has reached this goal.

In the Euclidean metric the general form of the matrix
element for semi-leptonic transitions $B \to B^\prime e\overline{\nu}_e$ is
\begin{eqnarray}
   \langle B^\prime (p^\prime) | V_{\alpha}(q) + A_{\alpha}(q) | B(p) \rangle 
      = \bar{u}_{B^\prime} (p^\prime) ( {\cal O}_{\alpha}^{V} (q) 
                           + {\cal O}_{\alpha}^{A}(q)) u_B(p) \,,
\end{eqnarray}
where
\begin{eqnarray}
   {\cal O}_{\alpha}^V(q) 
      = \gamma_{\alpha} f_1^{\R} (q^2) 
          + \sigma_{\alpha \beta} q_{\beta} \frac{f_2^{\R}(q^{2})}{M_B
          + M_{B^\prime}} + i q_{\alpha}\frac{f_3^{\R}(q^{2})}{M_B + M_{B^\prime}} \,,
\label{eq:OV}
\end{eqnarray}
and
\begin{eqnarray}
   {\cal O}_{\alpha}^A(q) 
      = \gamma_{\alpha} \gamma_5 g_{1}^{\R}(q^{2}) 
          + \sigma_{\alpha \beta} q_{\beta} \gamma_{5} 
                   \frac{g_{2}^{\R}(q^{2})}{M_B + M_{B^\prime}} 
          + i q_{\alpha} \gamma_{5} \frac{g_{3}^{\R}(q^{2})}{M_B + M_{B^\prime}} \,.
\label{eq:OA}
\end{eqnarray} 
(Note that in our definition we follow \cite{sasaki08a}, by symmetrising the mass terms appearing
in the denominator.) The form factors $f_{1}^{\R}$ (vector), $f_{2}^{\R}$ (weak magnetism)
and $f_{3}^{\R}$ (induced scalar) correspond to the vector component, while 
$g_{1}^{\R}$ (axial-vector), $g_{2}^{\R}$ (weak electricity) and $g_{3}^{\R}$
(induced pseudoscalar) correspond to the axial-vector component
of the current.

Our longer term aim is primarily to determine the CKM matrix element $|V_{us}|$
from the $\Delta S = 1$ semi-leptonic decays, \cite{gaillard84a,bender68a}
\begin{eqnarray}
   \Gamma = {G_F^2 \over 60\pi^2}(M_B - M_{B^\prime})^5
            (1-3\delta)|V_{us}|^2 |f_1^{\R}(0)|^2
            \left( 1 + 3\left|{g_1^{\R}(0) \over f_1^{\R}(0)}\right|^2 
                     + \ldots \right) \,,
\end{eqnarray}
where $G_F$ is the Fermi constant, $\delta = (M_B-M_{B^\prime})/(M_B+M_{B^\prime})$.
Hence for a determination of $|V_{us}|$ we require a knowledge of
the form factors $f_1^{\R}(q^2)$ and $g_1^{\R}(q^2)$ at zero $4$-momentum
transfer, $q^2 = 0$, together with a chiral extrapolation to the physical
point. (When at zero momentum transfer the form factors $f_{1}^{\R}$ and
$g_{1}^{\R}$ are simply the vector, $g_V^{\R}$, and axial-vector,
$g_A^{\R}$, coupling.) Although this is a complementary determination to
the more common kaon semi-leptonic decay determination, it is more
complicated, not least because it involves an axial form factor. 

However, as a first step, in this work we have restricted our calculations
to the specific case of zero $3$-momentum transfer. Thus from
eq.~(\ref{eq:OV}) for the vector case, we compute the linear combination
\begin{eqnarray}
   A^{\R}_{\overline{B}^\prime K^VB}(q^2_{\max}) 
      = f_{1}^{\R}(q^{2}_{\max}) 
          - \frac{M_B - M_{B^\prime}}{M_B + M_{B^\prime}}f_{3}^{\R}(q^{2}_{\max})
      \equiv f_{0}^{\R}(q^2_{\max}) \,.
\end{eqnarray}
Similarly for the axial vector case, from eq.~(\ref{eq:OA}) we have
the combination
\begin{eqnarray}
   A^{\R}_{\overline{B}^\prime K^AB}(q^2_{\max})
      = g_1^{\R}(q^{2}_{\max}) 
          - \frac{M_B - M_{B^\prime}}{M_B + M_{B^\prime}}g_{2}^{\R}(q^{2}_{\max})
      \equiv g_{0}^{\R}(q^2_{\max})  \,.
\end{eqnarray}
(The notation $f_{0}^{\R}$, $g_{0}^{\R}$ is customary for these
form factor combinations.) This is not enough to determine $f_1^{\R}$
and $g_1^{\R}$ at $q^2 = 0$ and at the physical point. (Form factors of
matrix elements are functions of both $q^2$ and $\delta m_l$.) So in order
to disentangle the form factors and to explore the effects on
$f_1^{\R}$, $g_1^{\R}$ of symmetry breaking it will be required,
in the future, to examine these form factors at various values of
transferred momenta so that they can be separated as discussed in
\cite{sasaki08a,sasaki12a,guadagnoli06a}.
Phenomenological analyses are given in e.g.\
\cite{cabibbo03a,mateu05a,yamanishi07a}. In particular \cite{yamanishi07a}
introduces a method similar to ours.
The Ademello--Gatto theorem actually complicates the determination:
we now have to find small second order $SU(3)$ flavour symmetry effects,
which appear here to be very small.

Note that this determination does not require an explicit determination
of the lattice renormalisation constants $Z_V$ and $Z_A$; as we are
interested in deviations from the $SU(3)$ flavour symmetric
value, it is sufficient to normalise the result either at the symmetric
point or (equivalently) with the averages $X_{F_V}$ and $X_{D_A}$, $X_{F_A}$.


 \section{Conclusions} 


We have taken the first steps in determining $SU(3)$ symmetry breaking
effects for matrix elements of all bilinear quark operators for the
baryon octet. The strategy of lattice simulations from a point on the
$SU(3)$ flavour symmetric along the path to the physical point keeping
the average quark mass constant is ideally suited to investigating
these effects. While of intrinsic interest themselves, of more
phenomenological interest is the determination of form factors
relevant to determination of the CKM matrix element $|V_{us}|$.
This requires more complicated momentum transfer computations,
and an investigation of $O(\delta m_l^2)$ effects, both of which we
are now embarking upon.


\section{Acknowledgements}
The numerical configuration generation was performed using the
BQCD lattice QCD program, \cite{nakamura10a}, on the IBM
BlueGeneL at EPCC (Edinburgh, UK), the BlueGeneL and P at
NIC (J\"ulich, Germany), the SGI ICE 8200 at
HLRN (Berlin--Hannover, Germany) and the JSCC (Moscow, Russia).
The BlueGene codes were optimised using Bagel, \cite{boyle09a}.
The Chroma software library, \cite{edwards04a},
was used in the data analysis. This investigation has been supported partly
by the DFG under contract SFB/TR 55 (Hadron Physics from Lattice QCD)
and by the EU grants 283286 (Hadron Physics3), 227431 (Hadron Physics2)
and 238353 (ITN STRONGnet). JMZ is supported by the Australian Research
Council grant FT100100005. We thank all funding agencies.



\end{document}